\documentclass[aps,prl,twocolumn,groupedaddress]{revtex4}

\usepackage{graphicx}
\usepackage{amssymb}
\usepackage{epstopdf}
\def\be{\begin{equation}}   \def\ee{\end{equation}}
\def\eq#1{{Eq~(\ref{#1})}}    \def\fig#1{{Fig.\ref{#1}}}

\begin{document}

\title{Chromosome Oscillations in Mitosis}
\author{Otger Camp\`as$^{\dag\ddag}$ \& Pierre Sens$^{\dag\odot}$}
\affiliation{$^{\dag}$ Institut Curie, UMR 168, 26 rue d'Ulm,
F-75248 Paris Cedex 05, France\\
$^{\ddag}$ Departament d'ECM, Universitat de Barcelona,
Avinguda Diagonal  647, E-08028 Barcelona, Spain\\
$^\odot$ Physico-Chimie Th\'eorique, UMR 7083, ESPCI  10 rue Vauquelin, F-75231 Paris Cedex 05 - France\\
Otger.Campas@curie.fr or pierre.sens@espci.fr}

\date{July 27, 2005}




\maketitle


\textbf{Successful cell division requires a tight regulation of chromosome motion via the activity of molecular motors. Many of the key players at the origin of the forces generating the movement have been identified, but their spatial and temporal organization remains elusive\cite{Salmon_Mitch}. The protein complex {\em Kinetochore} on the chromosome associates with microtubules emanating from one of the spindle poles and drives the chromosome toward the pole\cite{Sharp,Koshland}. {\em Chromokinesin} motors on the chromosome arms also interact with microtubules, ejecting the chromosome away from the pole\cite{Compton,Yajima}. In animal cells, a mono-oriented chromosome (associated to a single pole) periodically switches between phases of poleward and away from the pole movement\cite{Bajer, Salmon01, Salmon02}, a behavior tentatively explained so far by the existence of a complex switching mechanism within the kinetochore itself\cite{Salmon01,Mitchison, Riedermodel,Hunt}. Here we show that the interplay between the morphology of the mitotic spindle and the collective kinetics of chromokinesins can account for the highly non-linear periodic chromosome motion. Our analysis provides a natural explanation for the origin of chromosome directional instability and for the mechanism by which chromosomes feel their position in space.}\\

\vspace{-0.5cm}

The characterization of the forces leading to chromosome motion is one of the central questions in mitosis research\cite{Salmon_Mitch}. The oscillatory movement of mono-oriented chromosomes observed in prometaphase that persists during chromosome congression, metaphase and early anaphase\cite{Bajer, Salmon01, Salmon02, Compton}, is a signature of the forces acting on the chromosome. The poleward (P) force exerted by the kinetochore is thought to be due to cytoplasmic dyneins\cite{Sharp} and microtubule (MT) depolymerization in the kinetochore \cite{ Koshland}. Chromokinesin motors associated with the chromosome arms\cite{Tokai} move toward the plus end of MTs\cite{Yajima} and generate the force moving the chromosome away-from-the-pole (AP)\cite{Compton,kapoor}. All the available models for the oscillatory movement  of the chromosome (often referred to as ``kinetochore directional instability'' \cite{Salmon02}) have as a common feature that the kinetochore somehow controls the switching between P and AP phases \cite{Mitchison,Salmon02,Riedermodel,Hunt}. 
However, the collective behavior of molecular motors can give rise to
dynamical instabilities \cite{Prost02} which have been observed in biological and biomimetic systems \cite{Julicher,OJOJOJOJ}. Indeed, inhibiting
chromokinesin activity leads to the disappearance of the oscillations and to
the collapse of mono-oriented chromosomes onto the centrosome
\cite{Compton}. In order to precisely assess the role of chromokinesins, we
analyze the balance of forces on a mono-oriented chromosome (Fig.1a).

\begin{figure}[h]
\vspace{-0.5cm}
\centerline {\includegraphics[width=6.5cm]{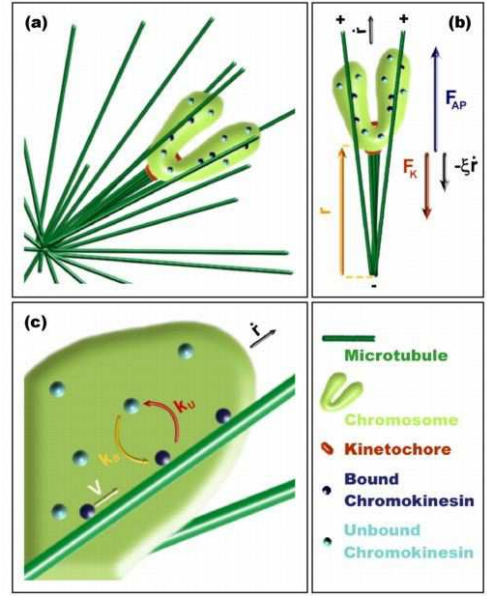}}
\caption{\label{fig1} \small Schematic representation of a mono-oriented
  chromosome. (a) Sketch of a MT aster (dark green) interacting with a single
  chromosome (light green). The kinetochore (red) is connected to the pole
  through a bundle of MTs. Chromokinesin motors on the chromosome arms may be
  bound to a MT (dark blue dots) or unbound (light blue dots).  (b) Forces
  driving chromosome motion: the kinetochore poleward force $F_K$ (red), the
  polar ejection force created by the bound chromokinesins $F_{AP}$ (dark
  blue) and the friction force opposing motion $-\xi \dot r$ (black). The chromosome postion relative to the pole (orange arrow) is $r$  and its velocity is $\dot r$. (c) Binding/unbinding kinetics of chromokinesin motors, with rates $k_b$ and $k_u$ respectively. In the bound state, chromokinesins move toward the plus end of MTs with a velocity $V$.}
\end{figure}

The chromosome motion occurs at length and velocity scales for which inertial
effects are negligible (low Reynolds number). As a result, a
difference between P and AP forces induces a viscous motion of the chromosome,
characterized by a phenomenological friction coefficient $\xi$. Force balance then reads
\be
F_{AP}-F_K-\xi\dot r=0
\label{balance}
\ee
where $\dot r\equiv dr/dt$ is the chromosome velocity ($t$ being the time),
$F_K$ is the P force applied by the kinetochore, and $F_{AP}$ the AP force
(Fig.1b). The friction coefficient $\xi$ generically characterizes the
resistance to chromosome motion, and pretends to encompass such effects as hydrodynamic friction, MT
polymerization and depolymerization and the activity of molecular motors in
the kinetochore\cite{Salmon03,Salmon_oscill}. Chromosome micromanipulation during anaphase
in grasshopper oocytes shows that the chromosome velocity $\dot r$ is
independent of chromosome position and decreases linearly with an external
force opposing the poleward motion\cite{Nicklas}. These observations strongly suggest that both the kinetochore force $F_K$ and the chromosome friction $\xi$ are nearly constant. We adopt such minimalist description of the kinetochore activity here and we estimate $F_K\sim 700pN$ and $\xi\sim 6\ 10^{-2}N.s/m$ from \cite{Nicklas}.

It has been shown that the kinetochore does not push on the chromosome, for instance through MT polymerization \cite{Salmon04,Riederpush}. Here we consider that the AP force is created solely by the binding and displacement of chromokinesin motors on the MT aster (Fig.1a).  Chromokinesins stochastically attach to and detach from MTs with average binding and unbinding rates $k_b$ and $k_u$ respectively. At a given time, only some amount $n(t)$ of the total number $N$ of chromokinesins permanently attached to the chromosome arms  is bound to MTs and able to participate to the AP force.  Generically, the time evolution of the number $n$ of bound motors may be written as
\be
\frac{dn}{dt} = k_b(N-n)-k_u n
\label{kineq1}
\ee

The higher the concentration of MTs, the easier it is for a motor to find an attachment site, hence the binding rate  $k_b$ is larger in dense regions of the MT aster. In a monopolar spindle (Fig.1a), the MT concentration decreases away from the pole, so the binding rate $k_b$ is a decreasing function of the chromosome position $r$. It is important to note that neither the P or the AP forces depend explicitly on chromosome position. All spatial information is contained in the binding rate $k_b(r)$ and reflects the morphology of  the MT spindle. 

Once bound, both the motor velocity $V$ and its unbinding rate are strongly
influenced by the motor load. If the $n$ bound chromokinesins are independent
from one another, they equally share the total ejection force $F_{AP}$ and each motor
feels a load $F_{AP}/n$. The rate of motor unbinding increases exponentially
with an applied load $k_u=k_u^{(0)}\exp[F_{AP}a/nk_BT]$ (Kramers
theory\cite{kramers}), where $k_u^{(0)}$ is the unbinding rate at vanishing
load, $a$ is a phenomenological length and $k_BT$ is the thermal energy. This
exponential sensitivity to applied force has been observed for conventional
kinesin \cite{Block03} (for which $k_u^{(0)}\simeq0.5s$\cite{Vale} and $a\simeq
1.3nm$\cite{Block03}) and also for myosin motors \cite{Veigel}. The velocity
of a motor decreases with a force opposing motor movement, and vanishes at
a particular stall force $f_s$. For conventional kinesin, the force-velocity relationship is nearly linear, with a velocity at vanishing force $V_0\simeq0.6\mu m/s$ and a stall force $f_s\simeq 6pN$\cite{Block02,Block}. We adopt the linear relationship $V=V_0(1-F_{AP}/nf_s)$, with no substantial influence on our results.

Identifying the chromosome velocity $\dot r$ with the chromokinesins velocity
$V$ on MTs and combining the equations above, we obtain a self-contained set
of two coupled equations for $\{r,n\}$ (see Supplementary Notes). The
origin of chromosome movement is fairly clear. Close to the pole, the MT
density is high, many chromokinesins attach and produce a large force that
moves the chromosome away from the pole. Far from the pole, MTs are scarce,
chromokinesins detach and the chromosome  moves poleward due to the
kinetochore force. Somewhere in between, there exists a fixed point where the
systems remains still. It corresponds to a number of bound chromokinesins
$n_s\equiv F_K/f_s$ that exactly balances the kinetochore force and to a
chromosome position $r_s$ (given by $k_b (r_s)= k_u^0 e^f n_s/(N- n_s)$)
where chromokinesin attachment and detachment fluxes exactly compensate (where $f\equiv f_s a/k_BT$ quantifies the influence of the motor load on its  detachment rate). 

The fixed point may be stable, in which case the chromosome remains at a fixed position $r_s$, or unstable, leading to permanent chromosome oscillations (see Supplementary Notes). Linear perturbation analysis around $\{r_s,n_s\}$ shows that the dynamical state of the chromosome is controlled by three parameters.  The dimensionless factor  $f\equiv f_s a/k_BT$ and the numbers $n_s=F_K/f_s$ and  $n_\xi\equiv \xi V_0/f_s$ normalized by the total number of chromokinesins $N$.
The fixed point is actually unstable for a wide range of parameters satisfying
$n_\xi<f n_s(1-n_s/N)-n_s$ (Fig.2). Note that the boundary between stable
and unstable regimes {\em does not} depend on the precise MT distribution in
the aster, provided the density decreases away from the pole (see Supplementary Notes). 
A reduction of the total number $N$ of available chromokinesins has the effect of moving the system along a straight line in the parameter space (arrow in Fig.2), and eventually to exit the oscillatory regime. The disappearance of chromosome oscillations has indeed been observed upon a drastic reduction of chromokinesins\cite{Compton}. We predict that the oscillation should stop for a precise number of chromokinesins: $N_c=f n_s^2/((f-1) n_s-n_\xi)$.

\begin{figure}
\centerline {\includegraphics[width=7cm]{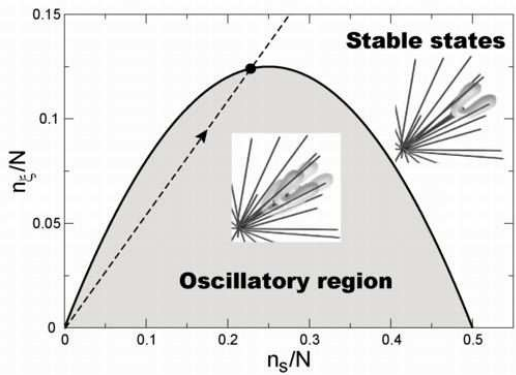}}
\caption{\label{fig2} \small
Dynamical regimes of chromosome motion. Parameter space representing the
region of stable chromosome position and chromosome oscillations ($f=2$). Increasing
$n_s$ and $n_\xi$ correspond to increasing kinetochore  force and chromosome
friction, respectively. Decreasing the total chromokinesin number $N$ (arrow)
eventually leads to the disappearance of oscillations at $N=N_c$ (filled circle).}
\end{figure}

Numerically computed chromosome motion in the unstable regime (Fig.3a) display the
characteristic sawtooth shaped oscillations observed {\em in-vivo}
\cite{Salmon02} (Fig.3c), indicating that the system switches suddenly
between phases of constant velocities.  Indeed, the period of the oscillation
($\sim min$) is controlled by the viscous motion of the chromosome, while the
switching between phases  occurs over much shorter time scales, characteristic of motor
binding/unbinding ($\sim sec$). Close to the pole, the high density of MTs
results in the binding of a large number of motors (Fig.3b), driving the chromosome AP
at a velocity close to their maximum velocity $V_0$. Thus, we argue that the
velocity of the AP motion is a direct  quantitative estimate of the
chromokinesin velocity at vanishing load ($V_0\simeq 2\mu$m/min in Newt lung
cells \cite{Salmon01,Salmon02}). As the chromosome moves away from the pole,
the density of MTs decreases and eventually reaches a value at which the attachment flux is too low to compensate the motor detachment. The remaining motors then detach rapidly (Fig.3b) and the chromosome  switches to P movement. The P phase occurs with almost no motors attached and the chromosome moves toward the pole with a constant velocity $-F_K/\xi$. The cycle is completed when the chromosome reaches a region of high enough MT density, where many motors abruptly attach and eject the chromosome. The ratio of AP and P velocities, which characterizes the symmetry of the oscillations, is  approximately given by $n_\xi/n_s$.   In Newt Lung cells (Fig.3c)\cite{Salmon02}, the poleward velocity is of order $2\mu$m/min, and the oscillations are roughly symmetrical. From their amplitude and period, we estimate $N\simeq 1000$ and $N_c\simeq 600$. Hence, the inactivation of about half the chromokinesins would be sufficient in this case to suppress oscillations.

\begin{figure}
\centerline {\includegraphics[width=9cm]{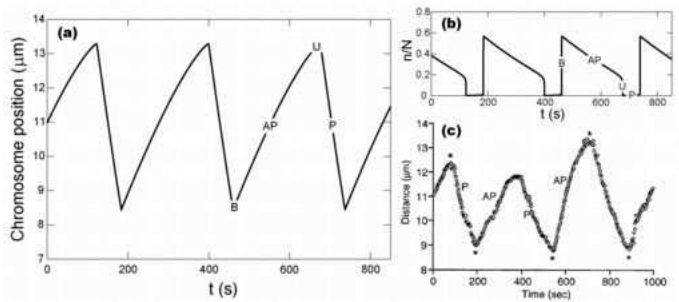}}
\vspace{-0.4cm}
\caption{\label{fig3} \small Chromosome oscillations. Position of the
  chromosome (a), and number of bound chromokinesins (b) as a function of
  time. The theoretical analysis reproduces and explains the sawtooth shape of
  {\em in-vivo} oscillations ((c) from \cite{Salmon_oscill}). The chromosome
  oscillates between phases of poleward (P) and away-from-the-pole (AP) motion
  with nearly constant velocities. The sudden switch from P to AP and AP to P
  phases corresponds to a dramatic binding (B) and unbinding (U) of
  chromokinesins respectively. The parameters are chosen to reproduce the
  amplitude and period of the {\em in-vivo} oscillations in (c). For the sake
  of simplicity, the MT distribution in the aster is assumed to be isotropic,
  so that $k_b(r)\sim1/r^2$. The AP velocity is an estimate of
  the chromokinesin velocity at vanishing load $V_0=2\mu m/min$ (see
  text). The remaining parameters are $k_u^{(0)}=1s^{-1}$, $k_b(r)=30/(r(\mu m))^2
  s^{-1}$), $n_s/N=0.115$, $n_\xi/N=0.052$ and $f=2$. }
\vspace{-0.5cm}
\end{figure}

We propose that chromokinesin binding onto MTs allows the  chromosome to feel its position in space via the aster-like morphology of the MT spindle and that chromokinesin unbinding from MTs is the force-sensitive mechanism at the origin of the chromosome instability. Within our framework, the highly non-linear, sawtooth shaped oscillations result from the combination of these two effects, and  not from a mechanism inside the kinetochore.  The kinetochore might also be force-sensitive, e.g. via the activity of cytoplasmic dyneins\cite{Salmon05}. Nevertheless, spatial information is essential for chromosome oscillations. We argue that it is provided by the position-dependent chromokinesin attachment rate and the spatial information provided by the MT density in the aster.

Finally, our analysis has implications for the congression of bi-oriented chromosomes\cite{kapoor}. The symmetry of a bipolar spindle insures that chromosomes correctly locate at the mitotic plate if the net force ($F_K-F_{AP}$) toward each pole increases with the distance to the pole.  Ostergreen\cite{ostergreen}  proposed that the kinetochore force $F_K$ increases linearly with the length of the MT fiber connecting it to the pole. We argue that the increase of the net poleward force is indeed nearly linear, but is rather due to decreasing polar ejection forces $F_{AP}$ away from the center of the MT aster (see Supplementary Discussion) . In conclusion, chromosome positioning may be due solely to the aster-like microtubule distribution and the kinetics of chromokinesins.

\begin{acknowledgments}
We thank J. Prost for critical reading of the manuscript and J-F. Joanny, F. J\"ulicher and J. Prost for stimulating remarks. O.C. thanks the European
Network PHYNECS (HPRN-CT-2006-00312) and Ministerio de Educaci\'on, Cultura y Deporte for financial support.\\
\end{acknowledgments}

\newpage
\renewcommand{\theequation}{SN-\arabic{equation}}
\renewcommand{\thesection}{SN-\arabic{section}}
\renewcommand{\thefigure}{SN-\arabic{figure}}
\setcounter{equation}{0}  
\setcounter{section}{0}  
\setcounter{figure}{0}  

\centerline{\Large Supplementary Notes}
\vspace{0.5cm}
{\bf{\em Dynamical regimes of chromosome movement}}

Mono-oriented chromosomes in a microtubule (MT) aster are subjected to poleward (P) force exerted by one of the chromosome {\em kinetochores} and to an away-from-the-pole (AP) force produced by {\em chromokinesin} motors on the chromosome arms.  The unbalance between these two forces leads to a viscous motion of the chromosome with a velocity $\dot r$ ($r$ is the distance from the pole). The number $n$ of chromokinesins exerting the AP force at one given time depends on  the kinetics of motor attachment/detachment to MTs, itself dependent of the morphological properties of the MT aster. The chromosome movement is fully determined by a set of coupled dynamical equations for $n$ and $r$, and the temporal evolution of the system corresponds to a given trajectory in the $\{n,r\}$ ``state space''. Using the equations given in the main text, the set of coupled differential equations (dynamical system) reads:
\begin{eqnarray}
\dot n&=&k_b(r)(N-n)-k_u^{(0)}\exp\left(f\frac{n_s+n_\xi}{n+n_\xi}\right)n \cr 
\dot r&=&V_0\frac{n-n_s}{n+n_\xi}
\label{kineq}
\end{eqnarray}
All spatial information is contained in the binding rate $k_b(r)$, which includes the MT density in the aster. The force-sensitive mechanism leading to the instability arises from the force-dependent detachment rate and the collective dynamics of motors: $k_u=k_u^{(0)}\exp (F_{AP} a/n k_BT)$. The ratio of the kinetochore force to the chromokinesin stall force $n_s=F_K/f_s$ is the number of bound motors necessary to compensate the kinetochore force, and the number $n_\xi=\xi V_0/f_s$ is the ratio of the chromosome friction to an effective chromokinesin friction.

This dynamical system has a single fixed point defined by $\dot n=0$ and $\dot r=0$, and given implicitly by  $\{n=n_s,r=r_s\}$ where
\begin{equation}
k_b(r_s)= k_u^0 e^f\frac{n_s}{N - n_s}
\label{fixedpoint}
\end{equation}
The linearized dynamics for the perturbations $\{\delta n,\delta r \}$ near the fixed point reads:
\begin{widetext}
\begin{eqnarray}
\frac{d}{dt}\left(
\begin{array}{c}\delta n \\ \delta r 
\end{array} \right) =
\underbrace{\left( \begin{array}{cc} 
k_u^0 e^f \left[ f \frac{n_s}{n_s+n_{\xi}}-\frac{N}{N-n_s} \right] &
\frac{\partial k_b(r_s)}{\partial r} \left( N -n_s\right) \\  \frac{V_0}{n_s+n_{\xi}} & 0 
\end{array} \right)}_{\mbox{$\Lambda$}} \left( 
\begin{array}{c} \delta n \\ \delta r 
\end{array}\right) 
\label{stability}
\end{eqnarray}
\end{widetext}

The matrix $\Lambda$ specifies the linearized dynamics of the system around the fixed point. The stability of the fixed point is obtained from the eigenvalues of $\Lambda$, or equivalently from its trace and determinant:
\begin{eqnarray}
\mbox{Det$(\Lambda)$}=-V_0 \frac{\partial k_b(r_s)}{\partial r} \frac{N -n_s}{n_s+n_{\xi}}\cr
\mbox{Tr$(\Lambda)$} = k_u^0 e^f \left[ -\frac{n_s}{N-n_s}+f\frac{n_s}{n_s+n_{\xi}}-1 \right]
\label{det}
\end{eqnarray} 
The determinant is always positive if enough motors are available ($N>n_s$) and for a decreasing concentration of MTs away from the pole ($\partial_r k_b(r)<0$). The stability of the fixed point is fully specified by the
trace of $\Lambda$. If Tr$(\Lambda)<0$, the fixed point is stable and any perturbation relaxes back to the fixed point. If
Tr$(\Lambda)>0$, perturbations are enhanced, driving the system away
from the fixed point. The equation  Tr$(\Lambda)=0$ sets the boundary between stable and unstable regimes. The stability of the systems is entirely determined by the three parameters: $n_s/N$, $n_\xi/N$ and $f$. The chromosome is unstable for $n_\xi<n_\xi^c=f n_s(1-n_s/N)-n_s$. The parameter $f$ quantifies the influence of the load on the motor detachment rate. If $f<1$, $n_\xi^c$ is always negative, implying that the fixed point is always stable. In other words, there exists a motor sensitivity threshold to force below which the chromosome always reach a stable position.

\begin{figure}[h]
\centerline {\includegraphics[width=8cm]{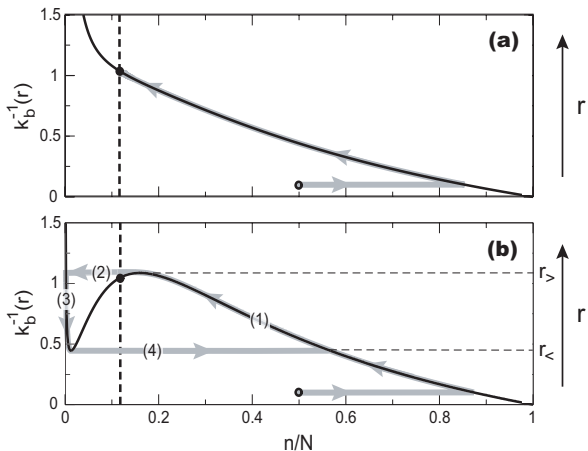}}
\vspace{-0.5cm}
\caption{\label{fig-SN} \small Stable and oscillatory trajectories in the ``State space'' of the dynamical system. Each point corresponds to a given number of bound motors $n$ and chromosome position $r$ ($k_b^{-1}(r)$, equivalently). The lines of vanishing chromosome velocity ($\dot r=0$ - dashed line) and of balance between motor binding and unbinding ($\dot n=0$ - black line) intersect at the fixed point (filled circle). The phase trajectories (thick gray line) start from an (arbitrary) initial condition (open circle). (a) Stable regime:  the n-nullcline is monotonous, and the systems evolves toward the (stable) fixed point. (b) Oscillatory regime:  the fixed point is unstable and the system follows a limit cycle corresponding to periodic oscillations between the extreme positions $r_<$ and $r_>$. Starting near the pole, the phases of the oscillatory chromosome movement are: (1) AP motion with many bound chromokinesins, (2) sudden motor unbinding far from the pole, (3) P motion with no bound motors and (4) sudden motor binding near the pole. Parameters: $ f=2$, $n_s=0.115 N$ and (for (a)) $n_\xi=0.152N$ (for (b)) $n_\xi=0.052N$.}
\end{figure}

In \fig{fig-SN} we represent the lines along which $\dot n=0$ ($n-$nullcline) and $\dot r=0$ ($r-$nullcline). The chromokinesin kinetics is typically much faster than the chromosome motion. As a consequence, the number of bound motor quickly adapts to a given chromosome position and the system always tries to follow the n-nullcline ($\dot n=0$). \fig{fig-SN}a shows a typical trajectory toward a stable fixed point. When the n-nullcline is non-monotonous (\fig{fig-SN}b), the region of positive slope of the n-nullcine is unstable because the detachment flux of bound motors is too high to be equilibrated by the attachment flux.  The linear stability analysis only specifies the transition between stable and unstable regimes. \fig{fig-SN}b shows that once the system becomes unstable, the trajectories in the ``state space'' $\{n,r\}$ evolve toward a limit cycle of the non-linear system, corresponding to permanent periodic chromosome oscillations. This can be shown formally using the Poincar\'e-Bendixon theorem (see J. Guckenheimer \& P. Holmes {\em Nonlinear Oscillations, Dynamical Systems, and Bifurcations of Vector Fields} (Applied Mathematical Sciences Vol. 42), Springer-Verlag, 1990). As can be seen in \fig{fig-SN}b, the unstable range for the number of bound motors corresponds to a range of chromosome positions ($r_<<r<r_>$) between which the chromosome oscillates.

\renewcommand{\theequation}{SD-\arabic{equation}}
\renewcommand{\thesection}{SD-\arabic{section}}
\renewcommand{\thefigure}{SD-\arabic{figure}}
\setcounter{equation}{0}  
\setcounter{section}{0}
\setcounter{figure}{0}  

\vspace{1cm}
\centerline{\Large Supplementary Discussion}
\vspace{0.5cm}
{\bf{\em ``Micromanipulation'' of mono and bi-oriented chromosomes}}

Molecular forces responsible for chromosome motion in mitosis may be probed by micromanipulation experiments where an external force  is applied to the chromosome. Here we investigate the effect of such force on the position of the chromosome within the mitotic spindle. 

\begin{figure}
\centerline {\includegraphics[width=8.6cm]{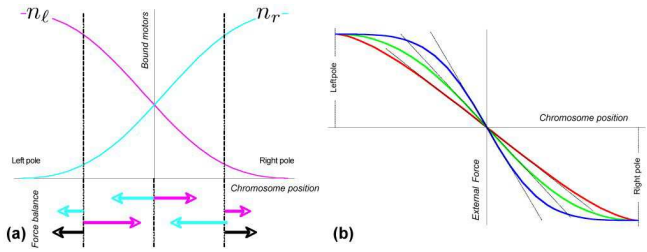}}
\caption{\label{fig-SD} \small (a) Number of chromokinesin motors bound to microtubules from the left (pink) and right (turquoise) asters. The balance of polar ejection forces and external force (black) is also represented  for two symmetrical chromosome positions. (b) 
Stable chromosome position as a function of the external force. The chromosome displacement with respect to the spindle mid-plane varies linearly with the force almost up to chromosome collapse onto one of the poles. Curve are for powerlaw decreases of microtubule density with distance from the pole $\alpha(r)=r^{-\beta}$ with $\beta=1.5$ (red) $\beta=2$ (green) and $\beta=3$ (blue)}
\end{figure}

The framework developed in the main text rely on the assumption of a constant force $F_K$ exerted by the kinetochore. If an additional external force $F_{ext}$ is applied to a mono-oriented chromosome in the direction of the pole, it merely shifts the effective value of the kinetochore force to $F_K+F_{ext}$ and modifies the chromosome motion accordingly. Bi-oriented chromosomes experience competing kinetochore forces from each pole and away-from-the-pole (AP) forces from both microtubule (MT) asters. The kinetochore forces being constant in our framework, they cancel out for each chromosome position. On the other hand, polar ejection forces implicitely depend on the chromosome position in the spindle via the chromokinesin binding rate and the local MT density. AP forces cancel at the mitotic plate due to the bipolar symmetry of the mitotic spindle (\fig{fig-SD}a). Note that bi-oriented chromosomes may also display the oscillatory behavior described in the main text. Here, we assume for the sake of clarity that a bi-oriented chromosome reaches a stable position, e.g. due to an increased friction from the bi-polar attachment. Chromokinesins on stable chromosomes always exert their stall force $f_s$ and the spatial dependence of the AP forces arises only from the variation of the number of bound motors. A chromosome under an external force $F_{ext}$ directed toward the left pole is shifted leftward to a position $r_s$ satisfying the force balance: 
\be
F_{ext}=(n_\ell(r_s)-n_r(r_s))f_s
\label{forcebal}
\ee
where $n_r$ and $n_\ell$ are the number of chromokinesins bound on MTs from the left and right asters, respectively.

In a monopolar spindle,  the spatial variation of the chromokinesin binding rate follows the spatial distribution of microtubules (MTs) in the aster and is given by the function $k_b(r)$. In a bipolar spindle with a distance $L$ between the spindle poles, we define the dimensionless function of local MT density  $\alpha(r)\equiv k_b(r)e^{-f}/k_u^{(0)}$, where $r$ is the distance from the left pole. At the stable position $r_s$, equilibration of motor attachment and detachment fluxes on each aster leads to:
\be
n_\ell(r)=n_r(L-r)=N\frac{\alpha(r)}{1+\alpha(r)+\alpha(L-r)}
\ee
where $N$ is the total number of available chromokinesins. Combining the two equations above gives a relationship between the applied force and the stable chromosome position:
\be
\frac{\alpha(r_s)-\alpha(L-r_s)}{1+\alpha(r_s)+\alpha(L-r_s)}=\frac{F_{ext}}{N f_s}
\label{eq}
\ee
 which could be directly measured in micromanipulation experiments.

 \eq{eq} is represented on \fig{fig-SD}b for various aster-like MT distributions. At vanishing external force, the stable position is at the mitotic plate $r_s=L/2$.  Due to the symmetry of the bipolar spindle, the chromosome displacement is linearly proportional to the external force even for fairly large displacements from the mid-plane. Mathematically, this arises from the existence of an inflexion point for the left-hand-side of \eq{eq}, so that the quadratic term in the Taylor expansion near the mid-plane vanishes. \fig{fig-SD}b shows that the linear relationship is valid almost up to values of the external force at which the chromosome collapses onto one centrosome: $F_{ext}\simeq N f_s$. According to our estimate, this force is in the nN range.

The situation we described has to some extent been realized experimentally  in (Hays, T.S., Wise, D. \& Salmon E.D. Traction force on a kinetochore at metaphase acts as a linear function of kinetochore fiber length {\em J. Cell Bio.} {\bf  93} 374-382  (1982)) by analyzing the metaphase stable position of multivalent chromosomes having more than two  kinetochores. In this case, the excess number of kinetochores generate an extra force toward one of the poles. The resulting chromosome displacement with respect to the mid-plane has indeed been reported  to be proportional to the excess number of kinetochores.

\end{document}